**Comment on "Neutron Lifetime and Axial Coupling Constant" by A. Czarnecki et al.**

Dirk Dubbers, Physikalisches Institut der Universität, INF 226, 69120, Heidelberg, Germany.
17 July 2018

*Abstract*: I give a comment on arXiv:1802.01804v3 [hep-ph].

Several new neutron lifetime and neutron $\beta$ asymmetry results were published in recent times. In this context, Czarnecki, Marciano, and Sirlin [1], outstanding theorists in the field, wrote a historical review on experimental neutron decay results. They primarily discuss the relation

$$\tau_n^{-1} = C\,(1 + 3g_A^2) \qquad (1)$$

between the lifetime $\tau_n$ and the axial coupling $g_A$ of the neutron. The ingredients of the prefactor $C = G_\mu^2\,|V_{ud}|^2\,(1 + RC)\,f\,m_e^5\,/\,2\pi^3 = (5172.0\,(1.1)\,\mathrm{s})^{-1}$ therein come from various sources and have not changed much over the years. Therefore relation (1) implies, in the authors' words, that "the $\tau_n$ and $g_A$ experimental values can be expected to move together". From this the authors come to their rather central conclusion that "the 2002 confirmation [11, 53] of a relatively large $g_A$ with small errors may be viewed as the harbinger of a shorter lifetime which several years later began to be directly observed in trapped lifetime experiments".

   I want to comment on this conclusion in order to prevent misinterpretation of the historical development of neutron decay experiments. I dislike to warm up old stories, but things were not as simple as seen from theory today: The authors' conclusion gives the impression that, based on our 2002 $\beta$ asymmetry result for $g_A$, it was clear as to what the results of ensuing neutron lifetime experiments should be. This conclusion, however, is contradicted by the fact that, up to 2010, the Particle Data Group [2] refused to include the then most precise 2005 neutron lifetime measurement (Ref. [19] in the article) with the argument that this lifetime "is so far from other results that it makes no sense to include it in the average", although the then adopted PDG 2010 values for $\tau_n$ and $g_A$ strongly violated relation (1). Furthermore, around 2005, there were problems with the unitarity of the CKM matrix, which caused considerable worry about the true value of $V_{ud}$ in (1), as was documented in several workshops, see for instance Ref. [3]. It was only several years later [4] that the CKM unitarity problem was shown to be caused not by $V_{ud}$, but by errors in the high-energy result for $V_{us}$. In contrast to this, the authors of [1] now give the impression that the 2005 lifetime result (others came much later) was obvious and as expected widely.